\documentclass[aps,prx,twocolumn,superscriptaddress,amsmath,amssymb,floatfix]{revtex4-1}
\usepackage{blindtext}
\usepackage{graphicx}
\usepackage{epsfig}
\usepackage{bm}
\usepackage{color}

\begin{document}

\title{Dislocation Majorana zero modes in perovskite oxide 2DEG}

\author{Suk Bum Chung}
\email{sbchung@snu.ac.kr}
\affiliation{Center for Correlated Electron Systems, Institute for Basic Science}
\affiliation{Department of Physics and Astronomy, Seoul National University, Seoul 151-747, Korea}
\author{Cheung Chan}
\email{phcchan@mail.tsinghua.edu.cn}
\affiliation{Institute for Advanced Study, Tsinghua University, Beijing 100084, China}
\author{Hong Yao}
\email{yaohong@tsinghua.edu.cn}
\affiliation{Institute for Advanced Study, Tsinghua University, Beijing 100084, China}

\date{\today}

\begin{abstract}
Much of the current experimental efforts for detecting Majorana zero modes have been centered on probing the boundary of quantum wires with strong spin-orbit coupling. The same type of Majorana zero mode can also be realized at crystalline dislocations in 2D superconductors with the nontrivial weak topological indices. Unlike at an Abrikosov vortex, at such a dislocation, there are no other low-lying midgap states than the Majorana zero mode so that it avoids usual complications encountered in experimental detections such as scanning tunneling microscope (STM) measurements. We will show that, using the anisotropic dispersion of the $t_{2g}$ orbitals of Ti or Ta atoms, such a weak topological superconductivity can be realized when the surface 2DEG of SrTiO$_3$ or KTaO$_3$ becomes superconducting, which can occur through either intrinsic pairing or proximity to existing s-wave superconductors.
\end{abstract}

\maketitle
\section{Introduction}
Non-Abelian braiding statistics of well-separated Majorana zero modes can provide one simpler means for realizing topological quantum computations \cite{Kitaev2003,Nayak2008}.
Partly motivated by this, the search for Majorana zero modes in nature has become one of the central and challenging issues in condensed matter physics in last few years \cite{Alicea2012,Franz2015}. It has been recognized in recent years that superconductivity in a system where the spin-orbit coupling and the Zeeman field co-exist can be topologically non-trivial even with conventional $s$-wave pairings, giving rise to Majorana zero modes in topological defects \cite{Sato2009, Sau2010, Alicea2010, Oreg2010}. Although much of the experimental investigations into this physics have focused on 1D systems, {\it e.g.} Refs \cite{Mourik2012, Nadj-Perge2014}, the results are known to hold in two-dimentional (2D) systems as well. One promising arena to realize such 2D systems is the two-dimensional electronic gases (2DEGs) formed at the boundaries of perovskite transition metal oxides \cite{Hwang2012}. Notable examples of such 2DEGs include the (001) interface between SrTiO$_3$ and LaAlO$_3$ \cite{Ohtomo2004}, the surfaces of SrTiO$_3$ \cite{Santander2011, Meevasana2011}, and KaTaO$_3$ \cite{King2012}. All of them possess nonzero Rashba-type of spin-orbit coupling because of the lack of inversion symmetry at the boundaries.  In addition, ferromagnetism is a frequent feature of these 2DEGs \cite{Brinkman2007, Ariando2011, Li2011, Bert2011} even though spin-orbit coupling usually tends to suppress spin alignment. Given that intrinsic superconductivity has been observed in many of these 2DEGs \cite{Reyren2007, Caviglia2008, Li2011, Bert2011, Kim2011, Ueno2011}, not to mention the possibility of inducing superconductivity on the surface 2DEGs through superconducting proximity effect, one naturally asks the question whether topologically protected Majorana zero modes can be achieved in the oxide 2DEG.

One major difficulty in investigating the topology of such oxide 2DEG superconductor is the multitude of bands at the Fermi level near the $\Gamma$ point. While the topologically nontrivial superconductivity in such 2DEGs has been studied with relatively simple models \cite{Fidkowski2013, Scheurer2015}, these studies rely on the assumption that only one or two bands crossing the Fermi level, which, however, seems to be at variance with the reported experimental data \cite{Caviglia2010}. This is because, unlike in the simplified models, the conduction bands of these 2DEGs near the $\Gamma$ point cannot be attributed to a single set of the transition metal $t_{2g}$ orbitals. Given that the formation of 2DEG requires a confining potential, it is likely that multiple quantum well channels arise from each $t_{2g}$ orbital \cite{Popovic2008}. Exactly how many bands occur at the $\Gamma$ point is often difficult to predict as the 2DEG confining potential is highly non-universal. Since the inclusion of more bands can turn topologically non-trivial superconductor into topologically trivial superconductor (and vice versa), this is an issue that raises question about the robustness of the simple model analysis.

We show here that it is possible to obtain robust Majorana zero modes in the perovskite oxide 2DEG by using its crystalline symmetry and anisotropic dispersions. Once the crystalline translational symmetry is assumed, weak topological indices are well-defined topological invariants \cite{Ran2009, Kitaev2009} that are unaffected by how many bands cross the Fermi level near the $\Gamma$ point. Physically, in the case of 2D superconductors breaking time-reversal-symmetry, weak topological indices can tell us whether there would be a Majorana zero mode on an edge dislocation \cite{Ran2010, Teo2010}, which can be detected with the STM tip. Although a Fermi surface crossing the boundary of the first Brillouin zone (BZ), a requirement for any non-trivial weak indices, has not been observed yet in the oxide 2DEGs, it is possible with currently available experimental techniques to tune the system to satisfy this condition. This is thanks to one universal feature of the (001) perovskite oxide 2DEG - the strong anisotropy of the band dispersion. This feature, signified by the sharp distinction between the light mass and heavy mass bands in the (001) perovskite oxide 2DEG ARPES (angle-resolved photoemission spectrum) data \cite{Santander2011, Meevasana2011, King2012}, means that there need not be large changes in either the Fermi level or the number of electrons per unit cell in tuning the system from the band bottom to the Lifshitz phase transition point. Moreover, this Lifshitz phase transition would involve only a single heavy mass band as all the light mass bands would be at much higher energy. Indeed, this tunability makes the oxide 2DEG a 
unique physical system to realize the dislocation Majorana zero mode compared to the ones discussed previously \cite{Asahi2012, Teo2013, Hughes2014, Benalcazar2014} as we shall show below.

\begin{figure*}[bht]
\centerline{\includegraphics[width=0.95\textwidth]{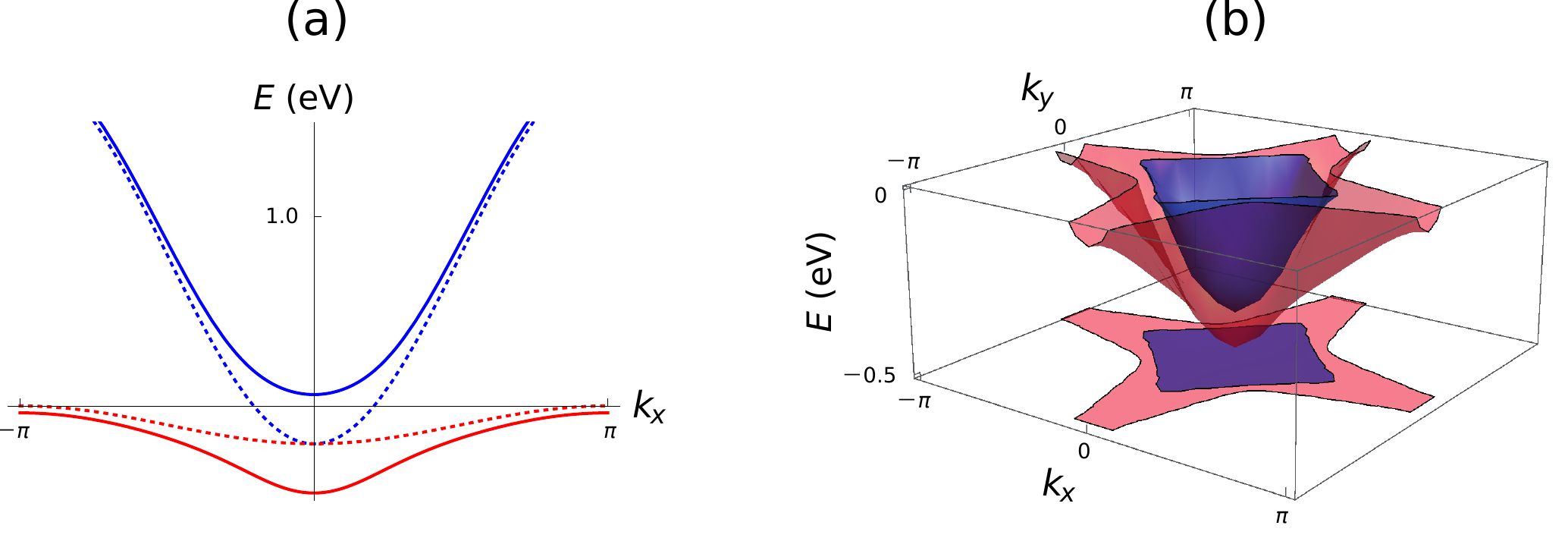}}
\caption{The band structure of the (001) perovskite oxide 2DEG near the Lifshitz transition. (a) shows the light- and heavy-mass band dispersion (in blue and red, respectively), along $k_y=\pi$ with (solid line) and without (dotted line) the orbital hybridization. (b) shows the lower band from the $d_{xz}/d_{yz}$ orbitals after the orbital hybridization, with the spin degeneracy removed by the Rashba spin orbit coupling $\alpha_0$ = 0.05eV and the perpendicular Zeeman field $h_Z$ = 0.05 eV.  Note that the Lifshitz transition point is split, allowing a single hole pocket without spin degeneracy around $(\pi, \pi)$. 
\label{FIG:band structure}}
\end{figure*}

\section{Results}

\subsection{Band structure of the (001) perovskite oxide 2DEG}
It has been known that electrons close to the Fermi level in typical (001) perovskite 2DEGs are mainly from $t_{2g}$ bands; the bands formed by $d_{xz}$ and $d_{yz}$ orbitals are quasi-1D while the one by $d_{xy}$ orbital is quasi-2D. As we are mainly interested in weak topological superconductivity in such systems, we shall focus on the electrons at the Brillouin zone boundaries, namely $k_x=\pi$ or $k_y=\pi$. One important feature of the (001) perovskite 2DEG band structure is that the electrons close to the Fermi level with $k_{x}=\pi$ ($k_y=\pi$) originate mostly from the quasi-1D $d_{yz}$ ($d_{xz}$) orbitals. This is because the low-energy physics at the BZ boundary arises out of the heavy-mass bands, and the quasi-1D nature of the $d_{xz} (d_{yz})$ orbital implies the heavy-mass dispersion in the $y(x)$-direction as well as the light-mass dispersion in the $x(y)$-direction. This anisotropic dispersion can be captured by the tight-binding model,
\begin{eqnarray}
\hat{K}_0 = &-& \sum_{{\bf r},\sigma}t\big[c^\dagger_{{\bf r}+{\bf \hat{e}}_x,x,\sigma}c_{{\bf r},x,\sigma} + c^\dagger_{{\bf r}+{\bf \hat{e}}_y,y,\sigma} c_{{\bf r},y,\sigma}+ {\rm h.c.}\big]\nonumber\\
&-&\sum_{{\bf r},\sigma}t'\big[c^\dagger_{{\bf r}+{\bf \hat{e}}_y,x,\sigma} c_{{\bf r},x,\sigma} +  c^\dagger_{{\bf r}+{\bf \hat{e}}_x,y,\sigma} c_{{\bf r},y,\sigma}+ {\rm h.c.}\big],~~
\end{eqnarray}
where $c^\dag_{{\bf r},a,\sigma}$ creates an electrons at site ${\bf r}$ with spin polarization $\sigma=\uparrow,\downarrow$ and orbital $a=x,y$ (representing $d_{xz}$ and $d_{yz}$ orbitals respectively). Because of the quasi-1D natures of $d_{xz}$ and $d_{yz}$ orbitals, $|t| \gg |t'|$. This simple model is sufficient to explain why we can reach the Lifshitz transition by lifting the Fermi level only in the order of the heavy-mass dispersion bandwidth $4t'$ and the electron filling fraction only by $\sim\sqrt{t'/t}$; this would involve raising the Fermi level by $\sim 0.1$eV and adding 0.3 electrons per unit cell when compared to the KTaO$_3$ surface ARPES data \cite{King2012} \footnote{We also note that superconductivity has been observed around this 2D electron density for the ionic liquid gated KTaO$_3$ \cite{Ueno2011}.}. Such shift in the Fermi level can be achieved by both the electrical gating and optically induced oxygen vacancies \cite{Meevasana2011, King2012, Ueno2012, Rice2014}. Meanwhile, the contribution from the $d_{xy}$ orbital is suppressed as it has the light-mass dispersion in both the $x$- and the $y$-direction.

We further need to consider the hybridization between the $d_{xz}$ and $d_{yz}$ orbitals in order to obtain from them two bands, one giving rise to the outer Fermi surface closer to the van Hove singularity at crystalline momentum points $X=(\pi,0)$  and $Y=(0,\pi)$ and the other giving rise to the inner Fermi surface closer to the $\Gamma$ point. Microscopically, the hybridization between $d_{xz}$ and $d_{yz}$ orbitals is mainly due to the on-site atomic spin-orbit coupling $\hat{K}_{aSO} = -\lambda \sum_{{\bf r}} c^\dagger_{{\bf r}} s_z\tau_y c_{{\bf r}}$ and the 
next-nearest neighbor hopping $\hat{K}_{nnn}=-t'' \sum_{{\bf r}}[(c^\dagger_{{\bf r}+{\bf \hat{e}}_x+{\bf \hat{e}}_y} s_0\tau_x c_{{\bf r}} - c^\dagger_{{\bf r}-{\bf \hat{e}}_x+{\bf \hat{e}}_y} s_0\tau_x c_{{\bf r}})+{\rm h.c.}]$, where $s_{\alpha}$ are Pauli matrices with spin indices and $\tau_\alpha$ with orbital indices. From these hybridization terms, we obtain the outer Fermi surface dispersion of
\begin{eqnarray}
&&\xi ({\bf k}) = -(t+t') (\cos k_x + \cos k_y) - \mu \\
&&- \sqrt{(t-t')^2 (\cos k_x - \cos k_y)^2 + \lambda^2 + (4t'' \sin k_x \sin k_y)^2}.\nonumber
\label{EQ:outerBand}
\end{eqnarray}
It is clear that, for $(t-t') \gg \lambda, t''$, the orbital hybridization would have little effect near the $X/Y$ points except for shifting the Lifshitz transition to $\mu = - \sqrt{4 (t-t')^2 +\lambda^2}$. Even with a large $\lambda$, as shown in Fig.\ref{FIG:band structure} (a) where we used $t = 10t' = 0.5$eV and $\lambda$ = 0.26eV, the former approximating the first principle calculation for KTaO$_3$ \cite{King2012, Shanavas2014, Kim2014} while the latter larger by roughly a factor of 2, the distinction between the heavy-mass and the light-mass bands remains sharp. Hence the physics near the $(\pi,0)$ point would be dominated by the $d_{yz}$ and near $(0,\pi)$ by the $d_{xz}$ orbital.

Since the inversion symmetry is obviously broken in the surface 2DEG, the spin degeneracy at the Fermi surface should be generically split by the non-zero Rashba spin-orbit coupling. For our analysis, it will be sufficient to consider only the most generic Rashba term, which is orbital independent,
\begin{equation}
\hat{K}^{(0)}_{RSO} = \alpha_0 \sum_{\bf r} \Big[c^\dagger_{{\bf r}+{\bf \hat{e}}_x} is_y\tau_0 c_{\bf r}-c^\dagger_{{\bf r}+{\bf \hat{e}}_y} is_x\tau_0 c_{\bf r} + {\rm h.c.}\Big].  
\label{EQ:L-vH}
\end{equation}
Further discussions on the spin-momentum coupling term will be presented in Supplementary Material \ref{App:SOC}.

After taking into account the hybridization as well as spin-orbital couplings, the band dispersions are described by $\hat K=\hat K_0+\hat K_{aSO}+\hat K_{nnn}+\hat K^{(0)}_{RSO}$. As the chemical potential $\mu$ moves, there is a Lifshitz transition at which the outer Fermi surface crosses the van Hove points at $X$ and $Y$. As we approach the Lifshitz transition, the low-energy band structure near the $(\pi,0)$ point, which would mainly originate from the $d_{yz}$ orbital, can be given by the first-quantized Hamiltonian
\begin{eqnarray}
\mathcal{H}_0(k_x, k_y) &=& -2t' \cos k_x - 2t \cos k_y -\mu \nonumber\\
&&- 2\alpha_0 [s_y \sin k_x - s_x  \sin k_y].
\label{EQ:vHband}
\end{eqnarray} 
Likewise the band structure in the vicinity of $(0,\pi)$, which would originate mainly from the $d_{xz}$ orbital can be obtained by the $\pi/2$ rotation of the momentum and the spin in Eq. \eqref{EQ:vHband}. It is also analogous to the Rashba wire that the spin degeneracy at $(0,\pi)$ remains unbroken, which means that the Fermi surface splitting does not lead to two separated Lifshitz transitions.

To possibly realize isolated Majorana zero modes, the final component needed for the band structure is the Zeeman field. Near the Lifshitz transition, there will be both higher density of oxygen vacancies near the surface as well as enhancement of the quasi-1D characteristics of the $d_{xz,yz}$ orbitals. Both can give rise to ferromagnetism: the former \cite{Rice2014, Michaeli2012} because of the oxygen vacancy acting as the magnetic impurity \cite{Lin2013} while the latter through the inter-orbital Hund's rule coupling \cite{Chen2013}. Both of these effects should be amplified by the enhanced density of states near the van Hove singularity that occurs at the Lifshitz transition. We will consider the ferromagnetic ordering in the perpendicular direction as was observed in the experiment with the density of oxygen vacancy \cite{Rice2014}. Then, the ferromagnetism-induced Zeeman coupling $\hat K_Z=h_Z\sum_{\bf r}c^\dag_{\bf r} s_z\tau_0 c_{\bf r}$ shall split the Lifshitz transition into two separated ones, as shown in Fig.\ref{FIG:band structure} (b), giving rise to a finite range of $\mu$ for which there is a single hole pocket without spin degeneracy around the $M=(\pi,\pi)$ point; for this plot we used $t'' = t'$ = 0.05eV  with $\alpha_0$ = 0.05eV and $h_Z$ = 0.05 eV. 

\begin{figure*}[bht]
\centerline{\includegraphics[width=0.95\textwidth]{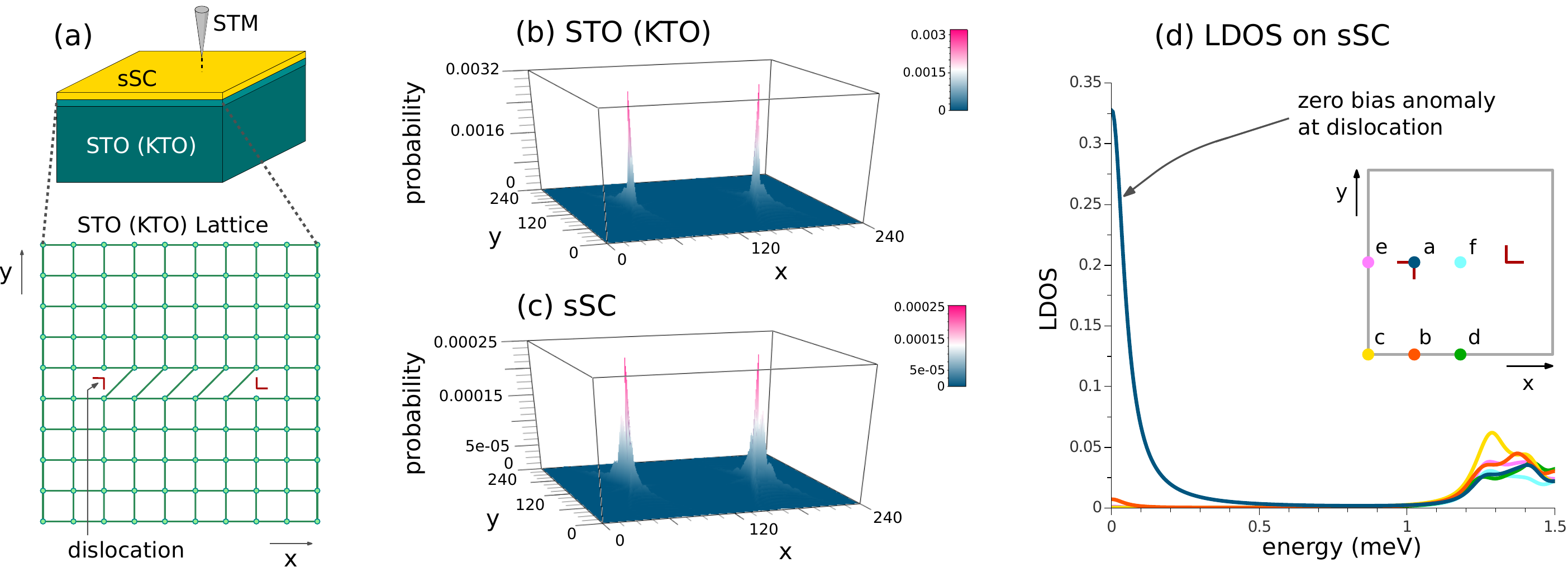}}
\caption{(a) shows the schematic setup of our system, with the oxide 2DEG superconductivity induced through the proximity effect and a pair of edge dislocations is on the oxide (STO or KTO) but not on the $s$-wave superconductor (sSC); an STM tip probes sSC. (b) and (c) shows the wave function profile for the Majorana zero modes on both the oxide surface and the $s$-wave superconductor. (d) plots the local density of states on the $s$-wave superconductor for various different points of the system, with the dark blue curve (point $a$) being taken right at the oxide dislocation position; note that both the sharp peak right above the oxide dislocation and the Majorana zero mode being the only subgap mode below the induced oxide pairing gap of $\sim$1.2 meV. \label{FIG:dislocation MZ}}
\end{figure*}

\subsection{Dislocation Majorana zero mode in proximity induced superconductivity}

For the superconducting state, we will first consider the case where the pairing is induced through proximity to the conventional $s$-wave superconductor. This will ensure the $s$-wave pairing in the oxide surface 2DEG. We also point out that inducing superconductivity through proximity effect can have the advantage of achieving superconductivity at higher temperature. To enhance the pairing gap magnitude, we would need strong tunneling between the superconductor and the oxide surface 2DEG. This can be achieved through using the higher-$T_c$ two-band superconductors such as FeSe \cite{QKXue2012,DLFeng2014,JFJia2014}; note that by symmetry, the single orbital superconductor is unlikely to have a strong tunneling to the both the $d_{xz}$ and $d_{yz}$ orbitals. Hence our heterostructure will consist of the capping two-band $s$-wave superconductor on the (001) surface of SrTiO$_3$ or KTaO$_3$ as shown in Fig.\ref{FIG:dislocation MZ} (a).

The combination of the Zeeman field $h_Z$ and the $s$-wave pairing gap $|\Delta_s|$ in the oxide 2DEG near the Lifshitz transition can give rise to the non-trivial weak index, ${\bm \nu} = (1,1)$, {\it i.e.} non-trivial 1D topological invariants along $k_{x,y} = \pi$. For instance, the following low-energy effective BdG Hamiltonian with $k_x=\pi$ is exactly equivalent to the Rashba-Zeeman wire superconducting state \cite{Oreg2010, Alicea2011}.
\begin{eqnarray}
&&\mathcal{H}_{sBdG} (k_x=\pi,k_y)\\
&=& \mu_z [2t (1-\cos k_y) - \delta\mu  + 2\alpha_0 s_x  \sin k_y]- s_z h_Z  + \mu_x |\Delta_s|, \nonumber
\label{EQ:1D}
\end{eqnarray}
where $\mu_\alpha$'s are Pauli matrices acting on the particle-hole Nambu space, $\delta \mu$ is the deviation of the chemical potential from the value at the Lifshitz transition for $h_Z = 0$, and we use the basis $(c_{{\bf k}\uparrow}, c_{{\bf k}\downarrow}, -c^\dagger_{-{\bf k},\downarrow}, c^\dagger_{-{\bf k},\uparrow})$. It is well known that this 1D BdG Hamiltonian is topologically equivalent to the Kitaev chain \cite{Kitaev2001} (class D \cite{Altland1997}) when  $h_Z^2 > |\Delta_s|^2 +  (\delta \mu)^2$. The $C_{4v}$ point group symmetry of the perovskite surface 2DEG dictates that  $\mathcal{H}_{sBdG} (k_x, k_y = \pi)$ should be topologically identical to $\mathcal{H}_{sBdG} (k_x=\pi,k_y)$.

The orbital hybridization will not affect the topological nature of the superconductivity as long as the $s$-wave pairing is intra-orbital. Given that the $s$-wave pairing has no spin dependence, we see that even if we take into account the band hybridization and write $\mathcal{H} (k_x=\pi, k_y)$ in the band basis, the Zeeman coupling and the $s$-wave pairing terms will remain unchanged, and hence so remain the condition for the topologically non-trivial superconductivity; we leave the detailed discussion to Supplementary Material.

Because of the nontrivial weak topological indices ${\bm \nu} = (1,1)$, unpaired Majorana zero modes occur at dislocations whose Burger's vector ${\bf B}$ in units of lattice spacings satisfies ${\bf B} \cdot {\bm \nu}=1$ (mod 2), where mod 2 is from the $Z_2$ nature of weak topological indices in class D \cite{Schnyder2008}. To confirm this, we have performed BdG calculations of the lattice models describing the 2DEG in proximity to a two-orbital $s$-wave superconductor (sSC). As shown in Fig.\ref{FIG:dislocation MZ} (b) and (c), we have obtained the Majorana zero mode at each dislocation from the numerical exact diagonalization of the real-space BdG Hamiltonian. Our calculation was done on a 240 $\times$ 240 unit cell with periodic boundary conditions. Two edge dislocations with the Burger's vector ${\bf B}=\pm \hat{\bf e}_x$ are placed by one half system size in the $x$-direction, with the links between the dislocations shifted as shown in Fig.\ref{FIG:dislocation MZ} (a) (see Methods for details on implementation). This oxide surface with the pair of dislocations is coupled by tunneling amplitude of $t_i =$ 0.05eV to the $s$-wave superconductor. The sSC has the band structure well-matched with that of the oxide surface (see Methods for the band structure details) and the pairing gap of  $|\Delta_s|$ = 0.05eV. Fig \ref{FIG:dislocation MZ} (b) and (c) shows the probability distribution of the dislocation zero energy states, showing sharp peak for both the oxide surface and the sSC, even though the latter does not have any dislocation. 

This wave function profile suggests that the STM would be a good experimental probe on our dislocation Majorana zero mode \cite{Law2009}. When the STM tip is brought to the sSC as shown in Fig. \ref{FIG:dislocation MZ} (a), the local differential conductance $\frac{dI}{dV}({\bf r},\omega)$ is proportional to the local density of state (LDOS) of the sSC, $\rho({\bf r}, \omega) = \sum_i [|u_i ({\bf r})|^2 \delta (\omega - E_i) + |v_i ({\bf r})|^2 \delta (\omega + E_i)]$ where the $u_i, v_i$ are the electron and hole components of the $i$-th energy eigenstates, up to replacing the delta function by a Lorentzian with the width given by the STM energy resolution, which is chosen to be 0.1meV for Fig.\ref{FIG:dislocation MZ} (d). We therefore predict that the STM will see a sharp zero bias anomaly when it is brought to the point on the sSC that is right over the dislocation, the point $a$ of Fig.\ref{FIG:dislocation MZ} (d). This anomaly is unambiguously separated from the signal of other low lying states, which has a minimum energy of the induced oxide bulk pairing gap $\sim$ 1.2meV. This is because the Majorana zero mode is the only midgap state localized at the dislocation, unlike at the Abrikosov vortex where other low energy ($\sim \frac{|\Delta|^2}{E_F}$, where $\Delta$ is the pairing gap and $E_F$ the Fermi energy) bound states are present. Hence the zero bias anomaly in the crystalline dislocation can be regarded as more unambiguous signature of the Majorana zero mode than that of the Abrikosov vortex \cite{Volovik1999}.

\subsection{Dislocation Majorana zero mode in intrinsic superconductivity}

\begin{figure*}[bht]
\centerline{\includegraphics[width=0.9\textwidth]{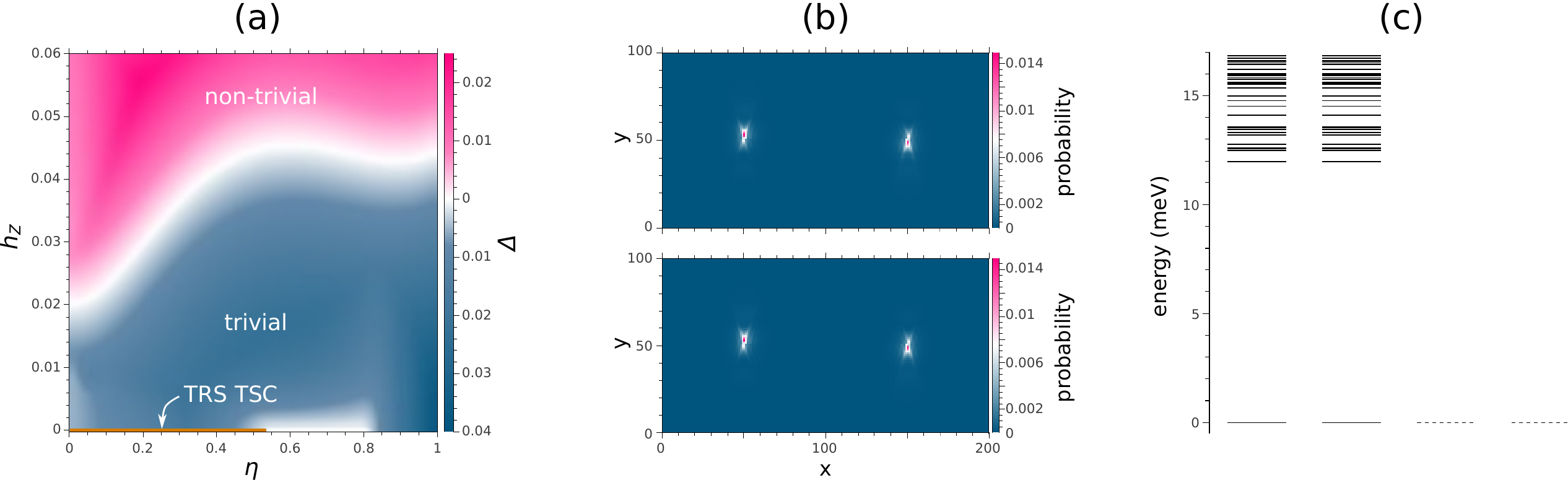}}
\caption{(a)  shows the energy gap $\Delta$ when we have mixture of the time-reversal invariant $p$-wave and the $s$-wave pairing, with $\eta = 0$ giving us the purely $p$-wave pairing and $\eta=1$ giving us the purely $s$-wave pairing. Note that for $h_Z \neq 0$, there need not be any gap closing in going from the pure $p$-wave to the pure $s$-wave pairing, suggesting that the two topological phases for the $p$-wave pairing - one at high $h_Z$ (in pink) and the other at low $h_Z$ (in blue) - are identical to those of the $s$-wave pairing. By contrast, for the case $h_Z=0$, where the time-reversal symmetry is preserved, the gap closing around $\eta = 0.7$ shows that the $p$-wave pairing (in orange) and the $s$-wave pairing are topologically distinct. (b)  shows the probability distributions for the two dislocation zero energy states obtained for $h_Z = 0$ and $\eta = 0.4$, which have exactly identical profile. (c) shows the energy level spacing for the eigenmodes with non-negative energies, again for $h_Z = 0$ and $\eta = 0.4$. For all positive energy eigenstates, separated from the zero energy by a gap of $\sim$12meV, there is double degeneracy due to the time-reversal symmetry. At the zero energy, we used the full and dotted lines to indicate the quadruple degeneracy at the zero energy from the occupancies and vacancies of the two zero energy states originating from the Kramer's double of Majorana zero modes at each dislocation. \label{FIG:intrinsic MZ}}
\end{figure*}

We now consider the case of intrinsic superconductivity in the oxide 2DEG without proximity to conventional superconductors. When the oxide 2DEG becomes superconducting at this electron density, there arises possibility of a protected Kramer's doublet of Majorana zero modes at each dislocation when no Zeeman field is applied. Due to the Rashba spin-orbit coupling, the intrinsic superconductivity should generically have on the Fermi surface a mixture of the $s$-wave pairing and the $p$-wave pairing, the latter with momentum-dependent Cooper pair spin state \footnote{This feature is independent of the debate on whether the pairing symmetry of the intrinsic superconductivity will follow \cite{Fidkowski2013} that of the doped bulk SrTiO$_3$ \cite{Schooley1964}, or not \cite{Scheurer2015}.}. For simplicity, we assume the following intra-orbital on-site and the nearest-neighbor pairing preserve the time-reversal symmetry \cite{Frigeri2004}:
\begin{eqnarray}
\hat{H}_{pair} &=& |\Delta_s|\sum_{\bf r} (c^\dagger_{{\bf r},\uparrow} \tau_0 c^\dagger_{{\bf r},\downarrow} + {\rm h.c.})\nonumber\\
&&-|\Delta_t|\sum_{\bf r}\sum_{i=x,y}[ c^\dagger_{{\bf r}+\hat{\bf e}_i} (s_i s_x) \tau_0 c^\dagger_{\bf r} + {\rm h.c.}]. 
\end{eqnarray}
With this pairing, the BdG Hamiltonian along the $k_x = \pi$ cut,
\begin{eqnarray}
\mathcal{H}_{tBdG} (k_x=\pi) &=& \mu_z [2t (1-\cos k_y) - \delta\mu  + 2\alpha_0 s_x  \sin k_y]\nonumber\\
&&+ \mu_x (|\Delta_s| - |\Delta_t| s_x\sin k_y),
\label{EQ:triplet}
\end{eqnarray}
will be that of a time-reversal invariant 1D topological superconductor in class DIII \cite{Schnyder2008,XLQi2009time} when $\delta \mu > 0$ and $|\Delta_s| < |\Delta_t| |\sin k_y|$ is satisfied at the Fermi surfaces so that the gaps at the two Fermi surfaces have the opposite signs. 
In that case, there exists of a branch of helical Majorana edge state around $k_{x,y} = \pi$. This means that, when we use the argument of the previous section with the additional constraint of the time-reversal symmetry, there should be a Kramer's doublet of Majorana zero modes at a dislocation with the Burger's vector of ${\bf B} = \hat{\bf x}$ or $\hat{\bf y}$. 

When the Zeeman field is non-zero, there can be a ``re-entrant'' unpaired Majorana zero modes at the dislocation. This is because Eq. \eqref{EQ:triplet} with the addition of the Zeeman field
\begin{eqnarray}
\mathcal{H}_{tBdG} (k_x=\pi, k_y) &=& \mu_z [2t (1-\cos k_y) - \delta\mu  + 2\alpha_0 s_x  \sin k_y]\nonumber\\
&-&s_z h_Z + \mu_x (|\Delta_s| - |\Delta_t| s_x\sin k_y),
\end{eqnarray}
in the $h_Z^2 > (\delta \mu)^2 + |\Delta_s|^2$ regime is topologically equivalent to the topological phase of Eq. \eqref{EQ:1D}. Hence, for $\delta \mu > 0$ at $h_Z = 0$, if the triplet pairing dominates, {\it i.e.} $|\Delta_s| < |\Delta_t| |\sin k_y|$ at the Fermi surfaces, there is a Kramer's doublet of Majorana zero modes at each dislocation for $h_Z = 0$ and a single Majorana zero mode for $h_Z > (\delta \mu)^2 + |\Delta_s|^2$.

Given that gap closing cannot be avoided at the topological quantum phase transition, Fig. \ref{FIG:intrinsic MZ} (a) gives us the complete topological phase diagram for the intrinsic superconductivity at a fixed $\delta \mu = 0.02$eV. In this plot, the tuning parameter $\eta$ is introduced to determine the relative strength of the $p$-wave and $s$-wave pairings, {\it i.e.} $\hat{\Delta} = \eta\hat{\Delta}_s + \tau_1 (1-\eta)|\Delta_{t0}|(s_y \sin k_x - s_x \sin k_y)$ where $\hat{\Delta}_s = |\Delta_{s0}| [\tau_1 + \tau_2 (2/\pi) (1-\eta) \arctan (h_Z/h_{Z0})]$ (we have set $|\Delta_{s0}| =$ 0.04eV, $|\Delta_{t0}| =$ 0.08eV and $h_{Z0} =$ 0.01 eV) \footnote{In order to have the pairing terms break the time-reversal symmetry when $h_Z \neq 0$, a phase difference between the $s$-wave and the $p$-wave pairings was introduced.}. For $h_Z \neq 0$, as shown in Fig. \ref{FIG:intrinsic MZ}(a), it is always possible to adiabatically tune $\eta$ from 0 to 1 without gap closing, while for any value of $\eta$, one cannot increase $h_Z$ without closing the bulk gap at some point. We therefore conclude that the non-trivial high (trivial low) $h_Z$ phase at $\eta = 1$ is topologically equivalent to that of $\eta = 0$, the purely $s$-wave pairing case we considered in the previous subsection. However, if we restrict ourselves to the case with time-reversal symmetry, {\it i.e.} $h_Z = 0$, Fig. \ref{FIG:intrinsic MZ} (a) shows there is a gap closing around $\eta = 0.7$, consistent with the fact that the time-reversal invariant $p$-wave pairing ($\eta = 0$) is topologically distinct from the pure $s$-wave pairing ($\eta = 1$). At $\eta = 0.4$ and $h_Z = 0$, the two zero energy states localized at the two edge dislocations with exactly identical probability distribution, as shown in Fig. \ref{FIG:intrinsic MZ}(b), indicates the existence of the Kramer's doublet of Majorana zero modes at each dislocation. This confirms that the existence of the Kramer's doublet of Majorana zero modes at each dislocation characterizes this time-reversal invariant topologically non-trivial phase.
This means that with an STM with an $s$-wave superconducting tip over this dislocation, we should be able to observe time-reversal anomaly \cite{Chung2013}. As in the case of the proximity induced superconductivity, these dislocation Majorana zero modes, as shown by Fig. \ref{FIG:intrinsic MZ}(c), are the only subgap modes of the system.

\section{Discussion}

We have shown in this paper how isolated dislocation Majorana zero mode can arise from both the proximity induced and intrinsic superconductivity in the oxide 2DEG. Its existence can be considered the most pertinent criterion for the topologically non-trivial superconductivity in the oxide 2DEG, and it can be experimentally detected through STM. The crucial requirement for achieving such superconductivity is that the oxide 2DEG needs to be close to the Lifshitz transition.

The key difference between 
the proximity-induced and the intrinsic oxide 2DEG superconductivity is that the Zeeman field is a necessary condition for the non-trivial topology in the former but not for the latter. The physical consequence is that for the  intrinsic superconductivity in the absence of the Zeeman field, the dislocation can host a Kramer's doublet of Majorana zero modes; this is not  possible if the superconductivity is induced through proximity to an $s$-wave superconductor. By contrast, in the presence of nonzero Zeeman field, the only possible protected midgap state on a dislocation is a single Majorana zero mode regardless of the origin of superconductivity.

While our intrinsic superconductivity with the non-trivial weak index at the zero Zeeman field has the essentially same pairing symmetry as the topological superconductivity investigated by Scheurer and Schmalian \cite{Scheurer2015}, these states are topologically distinct. From Eq.\eqref{EQ:triplet}, our non-trivial phase requires $\delta \mu >0$ while that of Scheurer and Schmalian requires $\delta \mu < 0$ with the Fermi surfaces enclosing the $\Gamma$ point, and with this pairing symmetry the gap closing around $\delta \mu = 0$ cannot be avoided. This reflects the fact that, with the reflection symmetry, the topological invariant of the DIII class in 2D can be $\mathbb{Z}$ rather than $\mathbb{Z}_2$ \cite{Chiu2013, Morimoto2013}. The existence (absence) of the dislocation Majorana doublet for $\delta \mu >0$ ($\delta \mu < 0$) can be regarded as a physical manifestation of this topological distinction. We leave to future work what type of interaction would favor this pairing symmetry near the Lifshitz transition.

Lastly, we want to point out that 
it is generically easy to change the topology of the superconducting state of the (001) perovskite oxide 2DEGs. This is because the universal anisotropic band structure makes it easy to access the van Hove singularity through gating and optically inducing oxygen vacancies. While there have been previous works on the physical realization of the the dislocation Majorana zero mode \cite{Asahi2012, Teo2013, Hughes2014, Benalcazar2014}, they have not provided easy means to alter the weak indices of the superconducting states. Therefore we conclude that not only is the dislocation Majorana zero mode the most robust topological feature of the oxide 2DEG superconductor but also that the oxide 2DEG superconductor is the particularly suitable system for realizing the dislocation Majorana zero mode.

\section{Methods}

\subsection{Weak indices and dislocation Majorana zero modes}

It is possible in a 2D superconductor on a square lattice to consider the 1D topological invariants defined along $k_{x,y} = \pi$, which are known as the weak indices \cite{Roy2009, Fu2007, Moore2007}.
In general, the weak index $\nu_i$ can be defined for each time-reversal invariant momentum ${\bf G}_i/2$ (which makes ${\bf G}_i$ a reciprocal lattice vector) as a topological invariant of the manifold perpendicular to ${\bf G}_i$ but contains ${\bf G}_i/2$, and hence the weak indices can be written as a single 
vector ${\bm \nu} = \sum_i \nu_i \hat{\bf G}_i$, where $\hat{\bf G}_i$ is the unit vector parallel to ${\bf G}_i$. The $C_{4v}$ symmetry of of our 2DEG means that its ${\bm \nu}$ will have only a single independent component $\nu$ and therefore can be written as ${\bm \nu} = \nu (1, 1)$. 
The weak indices is clearly topologically protected when the system has crystalline symmetry, the topological crystalline insulators\cite{Fu2011} being one class of examples. 
In this paper, we will focus on its manifestation through the Majorana zero mode localized at its crystalline topological defect - the edge dislocation \cite{Ran2010, Teo2010, Asahi2012, Teo2013}.

We first note that the non-trivial weak indices in a superconductor imply the existence of a branch of Majorana edge modes around $k_{edge} = \pi$. Since restricting ourselves to the $k_x=\pi$ manifold means converting the 2D mean-field Hamiltonian $H_{BdG} (k_x, k_y)$ into the 1D Hamiltonian $H_{BdG} (k_x = \pi, k_y)$, 
the non-trivial weak index means that, for the simplest case of the class D, where the time-reversal symmetry is broken, a single protected Majorana zero mode exists at $k_x = \pi$ for the edge running in the $x$-direction. This is possible only if there is a branch of chiral Majorana edge state centered around $k_x = \pi$. Note that the existence of this branch of the edge state is determined by the projection of ${\bm \nu}$ to the time-reversal invariant momentum $(\pi,0)$.

A single Majorana zero mode exists at the edge dislocation when there is a chiral Majorana edge state centered around $k_{edge} = \pi$. 
To see how this arises, note that the dislocation can be created by severing all links, both through hopping and interaction, between two halves ($y<0$ and $y>0$) and then non-trivially re-connect the two halves to introduce the edge dislocation, with the $x<0$ part glued back according to the original links but the $x>0$ part has all the links altered by translating the sites of the $y>0$ half by a lattice constant along the $x$-direction, which sets the Burger's vector of this dislocation to be ${\bf B} = {\bf \hat{x}}$ \cite{Ran2009, Ran2010}. Now when this system was cut, there would have been Majorana edge states along the $x$-direction for both $y<0$ and $y>0$ with the opposite chirality. Hence when the system is glued back along the original links, the tunneling between the two edges would lead to the backscattering that gaps out these edge modes, with the mass gap being proportional to the tunneling amplitude. However, when the dislocation described above is introduced, there will be a qualitative effect on the tunneling between the $k_x = \pi$ edge state. This is because the $k_x=\pi$ edge mode wave function reverses its sign when we translate by one lattice site along the $x$-direction, the relative sign of the $k_x=\pi$ edge modes for $y<0$ and $y>0$ edges will change its sign at the dislocation. That means that if a dislocation is introduced when we glue back with only infinitesimally weak coupling across $y=0$, the effective low energy action along $y=0$ for the  $k_x=\pi$ edge modes would be
\begin{equation}
S_{eff} = \int dx \Psi^T \left[\begin{array}{cc} i\partial_t + i\partial_x & im_0 {\rm sgn}(x)\\ -im_0 {\rm sgn}(x) & i\partial_t - i\partial_x \end{array}\right] \Psi
\end{equation}
where the upper and the lower component correspond to the upper and the lower edge and $m_0$ is proportional to the tunneling amplitude for the $k_x=\pi$ modes; this action is well-known for having a single Majorana zero mode at our dislocation $x=0$:
\begin{equation}
\Gamma_0 = \frac{\sqrt{m_0}}{2}\exp(-m_0 |x|)\left[\begin{array}{c} 1 \\ 1 \end{array}\right].
\end{equation}
By contrast, the existence of the $k_x=0$ branch is irrelevant as its tunneling amplitude does not change sign at the dislocation. Since the Majorana zero mode is protected as long as it remains separated from other Majorana zero mode, the Majorana zero mode that arose at the infinitesimal coupling across the $y=0$ cut will persist when the coupling across $y=0$ is increased to the bulk values. In general, the condition for the existence of the protected Majorana zero mode is ${\bm \nu} \cdot {\bf B} = 1$ (mod 2).

We can similarly show the existence of a Kramer's doublet of Majorana zero modes at the edge dislocation when there is a helical Majorana edge state centered around $k_{edge} = \pi$.  The key point here is that dislocation involves no time-reversal symmetry breaking and therefore, in the `cut and paste' picture, the Kramer's doublet needs to be maintained even with the inter-edge backscattering. Therefore, when $i \tilde{s}_y$ is the intra-edge time-reversal operation, the effective low energy action for the $k_x=\pi$ helical edge mode would be
\begin{eqnarray}
&&S_{eff} = \int dx 
\nonumber\\
&&\Psi^T\!\! \left[\!\!\begin{array}{cc} i\partial_t + i\tilde{s}_z\partial_x & i(\tilde{s}_z m + \tilde{s}_x m')  {\rm sgn}(x)\\ -i(\tilde{s}_z m + \tilde{s}_x m') {\rm sgn}(x) & i\partial_t - i\tilde{s}_z\partial_x \end{array}\!\!\right]\! \!\Psi.~~~~
\end{eqnarray}
This action gives us two Majorana zero modes,
\begin{eqnarray}
\Gamma_+ &=& \frac{\sqrt{m_0}}{2}\exp(-m |x|)\left[\begin{array}{c} \cos m' |x| \\ \sin m'|x| \\ \cos m'|x| \\ \sin m'|x| \end{array}\right],\\
\Gamma_- &=& \frac{\sqrt{m_0}}{2}\exp(-m |x|)\left[\begin{array}{c} -\sin m'|x| \\ \cos m'|x| \\ -\sin m'|x| \\ \cos m'|x| \end{array}\right],
\end{eqnarray}
which form a Kramer's doublet, {\it i.e.} $(i\tilde{s}_y)\Gamma_\pm = \pm \Gamma_\mp$.

\subsection{Real space Hamiltonian with dislocation}

We need to have the real-space BdG Hamiltonian in order to obtain the dislocation Majorana zero mode through exact diagonalization. We first note that the terms in our real-space Hamiltonian could be divided into three groups, the first being the onsite term,
\begin{align}
\hat{H}_{onsite} =& \sum_{\bf r} (-\mu + \sigma h_Z) c^\dagger_{{\bf r},a,\sigma} s_0 \tau_0  c_{{\bf r},a,\sigma}\nonumber\\
&- \lambda \sum_{\bf r}(c^\dagger_{\bf r} s_z \tau_y c_{\bf r} + {\rm h.c.})
+\hat{H}_{onsite-pair},
\end{align}
consisting of the chemical potential, the Zeeman energy, the atomic spin-orbit coupling and the onsite pairing, the second being the nearest neighbor  terms
\begin{align}
\hat{H}_{nn} =& \sum_{\bf r}\sum_{i=x,y} \hat{H}_{\hat{\bf e}_i} ({\bf r}),\nonumber\\
\hat{H}_{\hat{\bf e}_i} ({\bf r}) \equiv& -t \sum_{a=x,y} \delta_{a,i} (c^\dagger_{{\bf r}+\hat{\bf e}_i,a} s_0  c_{{\bf r},a} + {\rm h.c.})\nonumber\\
& -t' \sum_{a=x,y} (1-\delta_{a,i}) (c^\dagger_{{\bf r}+\hat{\bf e}_i,a} s_0  c_{{\bf r},a} + {\rm h.c.})\nonumber\\
& + \alpha \sum_a [c^\dagger_{{\bf r}+\hat{\bf e}_i,a}(i\epsilon_{ij} s_j) c_{{\bf r},a} + {\rm h.c.}]\nonumber\\
& +\hat{H}_{nn-pair}({\bf r}, \hat{\bf e}_i),
\end{align}
which includes the spin-conserving intra-orbital nearest neighbor hopping, the Rashba spin-orbit coupling, and the nearest-neighbor time-reversal invariant pairing. Lastly, we have the next-nearest neighbor hopping,
\begin{align}
\hat{H}_{nnn} =& \sum_{\bf r} \sum_{s=\pm}\hat{K}_{\hat{\bf e}_x+s\hat{\bf e}_y} ({\bf r}),\nonumber\\
\hat{K}_{\hat{\bf e}_x \pm \hat{\bf e}_y} ({\bf r}) \equiv& \mp t'' \sum_{\bf r}\sum_{s=\pm}(c^\dagger_{{\bf r}+{\bf \hat{e}}_x+s{\bf \hat{e}}_y} s_0 \tau_x c_{\bf r} +{\rm h.c.}).
\end{align}
which gives us the spin-independent component of the orbital hybridization. For the proximity-induced superconductivity, we set the nearest neighbor pairing to be zero, {\it i.e.} $\hat{H}_{nn-pair}({\bf r}, \hat{\bf e}_i)=0$, and set the onsite pairing to be originated entirely from a two-band $s$-wave superconductor:
\begin{eqnarray}
\hat{H}_{onsite-pair} &=& -t_i \sum_{\bf r} (c^\dagger_{\bf r} s_0 \tau_0  f_{\bf r} + {\rm h.c.}) + \hat{H}_s,
\end{eqnarray}
where $\hat{H_s}$ is given by
\begin{eqnarray}
\hat{H}_s &=& -t_s  \sum_{a=x,y} \delta_{a,i} (f^\dagger_{{\bf r}+\hat{\bf e}_i,a} s_0  f_{{\bf r},a} + {\rm h.c.})\nonumber\\
&& -t'_s \sum_{a=x,y} (1-\delta_{a,i}) (f^\dagger_{{\bf r}+\hat{\bf e}_i,a} s_0  f_{{\bf r},a} + {\rm h.c.})\nonumber\\
&-& \mu_s \sum_{\bf r}  f^\dagger_{\bf r} s_0 \tau_0  f_{\bf r}
+ |\Delta_s| \sum_{\bf r} (f^\dagger_{{\bf r},\uparrow} \tau_0 f^\dagger_{{\bf r},\downarrow} + {\rm h.c.})
\end{eqnarray}
(in Fig.\ref{FIG:dislocation MZ}, we have set $t_s=t, t'_s = t'$ and $\mu_s = -0.7$eV for the maximal proximity effect), while for the intrinsic superconductivity calculation shown in Fig.\ref{FIG:intrinsic MZ}, we set the pairing terms to be
\begin{align}
\hat{H}_{onsite-pair} =& \eta|\Delta_{s0}|[1-i(2/\pi)(1-\eta)\arctan(h_Z/h_{Z0})]\nonumber\\
&\times\sum_{\bf r} (c^\dagger_{{\bf r},\uparrow} \tau_0 c^\dagger_{{\bf r},\downarrow} + {\rm h.c.}),\nonumber\\
\hat{H}_{nn-pair}({\bf r}, \hat{\bf e}_i) =& -(1-\eta)|\Delta_{t0}| [c^\dagger_{{\bf r}+\hat{\bf e}_i} (s_i s_x) \tau_0 c^\dagger_{\bf r} +{\rm h.c.}],
\label{EQ:realSpacePair}
\end{align}
where we have set $|\Delta_{s0}| =$0.04eV, $|\Delta_{t0}| =$0.08eV, $h_{Z0} =$0.01eV with $0<\eta<1$ determining the relative contribution of the $p$- and $s$-wave pairings. For the intrinsic superconductivity under a finite $h_Z$, the phase difference between the $s$-wave and the $p$-wave pairing were added so that the pairing terms would break time-reversal symmetry.

In the real space, the dislocation point serves as a starting point for a branch cut along which the nearest-neighbor Hamiltonian $\hat{H}_{\hat{\bf e}_i} ({\bf r})$ 
is applied on a next-nearest neighbor link. Our square lattice $N_x = 240$ by $N_y =240$ latitce has a periodic boundary condition to both direction. In order to have a dislocations at $(N_x/4, N_y/2 + 1)$ with the Burger's vector ${\bf B} =+\hat{\bf e}_x$ and another dislocation at $(3N_x/4, N_y/2)$ with ${\bf B} =-\hat{\bf e}_x$, we apply $\hat{H}_{\hat{\bf e}_y}$ on the links connecting $(n, N_y/2)$ and $(n+1, N_y/2+1)$ for $N_x/4 \leq n < 3N_x/4$, $\hat{K}_{\hat{\bf e}_x + \hat{\bf e}_y}$ on the link connecting $(n, N_y/2)$ and $(n+2, N_y/2 + 1)$ for $N_x/4 \leq n < 3N_x/4-1$, and $\hat{K}_{\hat{\bf e}_x - \hat{\bf e}_y}$ on the link connecting $(n,N_y/2)$ and $(n, N_y/2 + 1)$; meanwhile between the two nearest neighbor pairs $(N_x/4, N_y/2)$ and $(N_x/4, N_y/2+1)$, $(3N_x/4, N_y/2)$ and $(3N_x/4, N_y/2+1)$ and also between the two next-nearest neighbor pairs $(N_x/4, N_y/2 + 1)$ and $(N_x/4+1, N_y/2)$, $(3N_x/4 - 1, N_y/2 + 1)$ and $(3N_x/4, N_y/2)$, all hoppings and pairings are set to zero. Note that for the case of proximity-induced superconductivity, the $s$-wave superconductor remains completely free of crystalline defects.

While we set some of the parameters to be rather large for the sake of convenience in the numerical calculation, such choice does not affect the topological properties of the system. For instance, $\lambda = 0.26$eV is about factor of 2 larger than the estimated value for the tantalum atom, while $\alpha_0 = 0.05$eV is several times larger than the estimated value from the magnetoconductivity measurement \cite{Caviglia2010a}. These choices are intended to increase the bulk energy gap so that our lattice size is sufficient to see a localized dislocation zero mode. This increase in the bulk energy gap occurs away from the BZ boundary $k_{x,y} = \pi$, {\it e.g.} the larger $\alpha_0$ increases the energy gap along $k_x = \pm k_y$, while the larger $\lambda$ lifts the higher $d_{xz/yz}$ band away from the Fermi level. Such changes do not affect weak indices, which are the 1D topological invariant along $k_{x,y} = \pi$. Concerning real materials and experiments, as long as the system is in the topological regime and the induced bulk gap is large enough, {\it i.e.} much larger than the STM resolution, the dislocation Majorana zero mode and the zero energy anomaly can be detected clearly as shown in Fig.\ref{FIG:dislocation MZ}(c).


Lastly, we point out that with our $p$-wave pairing in Fig. \ref{FIG:intrinsic MZ} (a) for $h_Z = 0$ allows for a finite range of $\eta$ for which there are nodal quasiparticles. While it is possible in principle to come up with a $p$-wave pairing for which the energy gap closes for only a single value of $\eta$, such $p$-wave pairing should have constant magnitude over the entire Fermi surface, which in general is not possible with our nearest-neighbor pairing.

{\it Acknowledgement}: We would like to thank Erez Berg, Harold Y. Hwang, Minu Kim, Takahiro Morimoto, Naoto Nagaosa, Tae Won Noh, Srinivas Raghu and Shoucheng Zhang for helpful discussions. This work is supported in part by IBS-R009-Y1 (SBC), the Thousand-Young-Talent Program of China (HY), and by the NSFC under Grant No. 11474175 (CC and HY).

\bibliography{oxide}

\begin{widetext}
\section*{Supplementary Material}
\renewcommand{\theequation}{S\arabic{equation}}
\setcounter{equation}{0}
\renewcommand{\thefigure}{S\arabic{figure}}
\setcounter{figure}{0}

\subsection*{A. Rashba spin-orbit coupling near van Hove singularity}
\label{App:SOC}


The $C_{4v}$ point group symmetry of the (001) perovskite surface 2DEG allows for the spin-momentum coupling in the $d_{xz,yz}$ orbitals from the following nearest-neighbor hopping terms\cite{Zhong2013,Scheurer2015}:
\begin{eqnarray}
\hat{K}_{RSO} &=& \alpha_0 \sum_{\bf r} \Big[c^\dagger_{{\bf r}+{\bf \hat{e}}_x} is_y\tau_0 c_{\bf r}-c^\dagger_{{\bf r}+{\bf \hat{e}}_y} is_x\tau_0 c_{\bf r} + {\rm h.c.}\Big]\nonumber\\
&+&\alpha_1  \sum_{\bf r}\Big[ c^\dagger_{{\bf r}+{\bf \hat{e}}_x} is_x\tau_x c_{\bf r}
-c^\dagger_{{\bf r}+{\bf \hat{e}}_y} is_y\tau_x c_{\bf r} + {\rm h.c.}\Big]+\alpha_3  \sum_{\bf r}\Big[ c^\dagger_{{\bf r}+{\bf \hat{e}}_x} is_y\tau_z c_{\bf r}
+c^\dagger_{{\bf r}+{\bf \hat{e}}_y} is_x\tau_z c_{\bf r} + {\rm h.c.}\Big].
\label{EQ:fullRashba}
\end{eqnarray}
This raises the question how these terms arise microscopically. Also we can ask whether there is any qualitative effects due to the orbital dependent terms $\alpha_{1,3}$, {\it i.e.} whether there is any $\sin k_{x,y}$-dependent spin-momentum locking of magnitude in the order of $\alpha_{0,1,3}$ along $k_{x,y} = \pi$.

The starting point for the microscopic physics is that due to the breaking of the inversion symmetry with respect to the $xy$-plane, there can be spin-independent hopping between $d_{xz,yz}$ orbitals and other $d$-orbitals:
\begin{align}
\hat{K}_{inv} =& -\gamma_1 \sum_{\bf r}\sum_{a,a'} (1-\delta_{a,a'})\Big[c^\dagger_{{\bf r}+{\bf \hat{e}}_a,a'} s_0 c'_{\bf r} - c'^\dagger_{{\bf r}+{\bf \hat{e}}_a} s_0 c_{{\bf r},a'} + {\rm h.c.}\Big] - \gamma_2 \sum_{\bf r}\sum_{a,a'} \delta_{a,a'}\Big[c^\dagger_{{\bf r}+{\bf \hat{e}}_a,a'} s_0 \tilde{c}_{\bf r} - \tilde{c}^\dagger_{{\bf r}+{\bf \hat{e}}_a} s_0 c_{{\bf r},a'} + {\rm h.c.}\Big]\nonumber\\
&-\gamma_3 \sum_{\bf r} \Big[c^\dagger_{{\bf r}+{\bf \hat{e}}_x,x} s_0 \tilde{c}'_{\bf r} - \tilde{c}'^\dagger_{{\bf r}+{\bf \hat{e}}_x} s_0 c_{{\bf r},x} - c^\dagger_{{\bf r}+{\bf \hat{e}}_y,y} s_0 \tilde{c}'_{\bf r} + \tilde{c}'^\dagger_{{\bf r}+{\bf \hat{e}}_x} s_0 c_{{\bf r},x} + {\rm h.c.}\Big],
\end{align}
where $c', \tilde{c}, \tilde{c}'$ are the annihilation operators for $d_{xy}, d_{z^2}, d_{x^2-y^2}$ orbitals respectively. Note that, in the bulk, any spin-conserving hybridization of this type would not be allowed due the fact that these orbitals are even under the inversion with respect to the $xy$-plane while the $d_{xz,yz}$ orbitals are odd. But even in the bulk, this hybridization is allowed through the transition metal spin-orbit coupling:
\begin{align}
\hat{K}'_{aSO} =& -\lambda \sum_{\bf r} \Big[c^\dagger_{{\bf r},x} (-is_x) c'_{\bf r} + c^\dagger_{{\bf r},y} (is_y) c'_{\bf r} + {\rm h.c.}\Big]\nonumber\\
& - \sqrt{3}\lambda \sum_{\bf r} \Big[c^\dagger_{{\bf r},x} (-is_y) \tilde{c}_{\bf r} + c^\dagger_{{\bf r},y} (is_x) \tilde{c}_{\bf r} + {\rm h.c.}\Big] - \lambda \sum_{\bf r} \Big[c^\dagger_{{\bf r},x} (is_y) \tilde{c}_{\bf r} + c^\dagger_{{\bf r},y} (is_x) \tilde{c}_{\bf r} + {\rm h.c.}\Big].
\end{align}
The hybridization of the $d_{xz,yz}$ orbitals with the other $d$-orbitals through $\hat{K}_{inv}+\hat{K}'_{aSO}$ effectively generates the Rashba-Dresselhaus terms of Eq.\eqref{EQ:fullRashba} in the subspace of $d_{xz,yz}$ orbitals \cite{Zhong2013, Kim2013, Scheurer2015}. One relatively simple way to obtain these Rashba-Dresselhaus terms is through the second order degenerate perturbation theory
\begin{equation}
\langle \tilde{a}, \sigma | \hat{\mathcal K}_{RSO} | \tilde{a}', \sigma' \rangle = \sum_{b \neq x,y} \frac{\langle \tilde{a}, \sigma |\hat{K}_{inv}|b, \sigma''\rangle \langle b,\sigma''|\hat{K}'_{aSO}|\tilde{a}', \sigma' \rangle + \langle \tilde{a}, \sigma |\hat{K}'_{aSO}|b, \sigma''\rangle \langle b,\sigma''|\hat{K}_{inv}|\tilde{a}', \sigma' \rangle}{\xi_0 - \xi_b},
\label{EQ:degRashba}
\end{equation}
the point here being that we take $\tilde{a}, \tilde{a}'$ to be the band (with spin degeneracy) formed by $d_{xz,yz}$.

One question that arises here is whether the orbital $b$ in Eq.\eqref{EQ:degRashba} should include the $e_g$ orbitals as well. Much of the analysis for the Rashba-Dresselhaus term near the $\Gamma$ point excludes the $e_g$ contributions to Eq.\eqref{EQ:degRashba} \cite{Zhong2013, Kim2013, Kim2014, Scheurer2015}, which would be justified in the limit where the $e_g$ orbital energies are much higher than that of the $d_{xz,yz}$ orbitals.
Applicability of this limit is qualitatively important for the analysis near the X point, for excluding the $e_g$ contribution effectively gives us 
\begin{equation}
\hat{\mathcal K}_{RSO} = \frac{4\gamma_1 \lambda}{\delta \xi_{xy,yz}} s_y \sin k_x,
\end{equation}
where $\delta \xi_{xy,yz}$ is the energy splitting between the $d_{xy}$ and $d_{yz}$ orbitals at the X point, for the $d_{yz}$ orbital (it is straightforward to obtain analogous result for the $d_{xz}$ orbital near the Y point), in which case it would not be justifiable to take the spin-momentum coupling to be $\hat{K}^{(0)}_{RSO}$ of Eq.\eqref{EQ:L-vH}. However, when the inversion symmetry is broken, the energy splitting between the two $e_g$ orbitals actually may be in the same order of magnitude as the crystal field splitting between the $t_{2g}$ and $e_g$ orbitals \cite{Shanavas2014, Shanavas2014a}. In such case, the contribution of the lower $e_g$ band (mostly from the $d_{z^2}$ orbital) to Eq.\eqref{EQ:degRashba} would give us \cite{Kim2015} 
\begin{equation}
\hat{\mathcal K}_{RSO} \approx \frac{4\gamma_1 \lambda}{\delta \xi_{xy,yz}} s_y \sin k_x + \frac{4\sqrt{3}\gamma_2 \lambda}{\delta \xi_{z^2,yz}} s_x \sin k_y. 
\end{equation}
Hence we are justified qualitatively in setting $\hat{K}^{(0)}_{RSO}$ of Eq.\eqref{EQ:L-vH} to be the spin-momentum coupling in the $d_{xz,yz}$ orbitals \footnote{We note that the inclusion of the $d_{z^2}$ orbital effect near the $\Gamma$ point insures that the lowest order Rashba term will be linear rather than cubic.}.

\subsection*{B. Orbital mixing effect near van Hove singularity}
\label{App:bandOrbital}

In this Section, we will show that even with a strong orbital hybridization, Eq.\eqref{EQ:1D} still gives us the effective low-energy Hamiltonian at $k_x = \pi$.

As a starting point, let us consider the first-quantized Hamiltonian in the normal state for the $d_{xz} / d_{yz}$ orbitals,
\begin{align}
\mathcal{H}_{normal}^{(0)} =& -(t+t')(\cos k_x + \cos k_y) - \mu
- \tau_z (t-t')(\cos k_x - \cos k_y)
+ \tau_y s_z \lambda + \tau_x 4 t'' \sin k_x \sin k_y
\end{align}
(where $\sigma_i$'s are the orbital Pauli matrices, with the $d_{xz/yz}$ being the eigenstate of $\sigma_3$ with the eigenvalue of $\pm 1$), where we have left out the Zeeman and the Rashba terms. When $\lambda$ is in the same order of magnitude as $t-t'$, as is known for KTaO$_3$, there would be a considerable orbital hybridization even when $k_{x,y} = \pi$. Along $k_x = \pi$, we can relate the band basis and orbital basis through the transformation $U = \exp(i\sigma_1 s_z \beta_{k_y}/2)$ where $\tan \beta_{k_y} = -\lambda/[(t-t')(1 + \cos k_y)]$, giving us
\begin{eqnarray}
U \left. \mathcal{H}_{normal}^{(0)} \right\vert_{k_x = \pi} U^\dagger &=& (t+t')(1-\cos k_y) - \mu
+\tau_z \sqrt{(t-t')^2 (1+\cos k_y)^2 + \lambda^2}\nonumber\\
&\approx&  (t+t')(1-\cos k_y) - \mu
+\tau_z \left[\sqrt{4(t-t')^2 + \lambda^2} -\frac{2(t-t')^2(1-\cos k_y)}{\sqrt{4(t-t')^2 + \lambda^2}}\right].
\end{eqnarray}
The low-energy Hamiltonian will then be given by the projection to the lower band, $\sigma_3 = -1$.

We then need to consider how the other terms, which are much smaller in the magnitude, transforms under $U$. Since it is obvious that the Zeeman term $-s_z h_Z$ and the intra-orbital $s$-wave pairing $\tau_1 |\Delta_s|$ remains invariant, we can mainly focus on the spin-momentum coupled nearest neighbor hopping. 
From Eq.\eqref{EQ:fullRashba}, we can see that these terms in the first-quantized form comes out to be
\begin{align}
\mathcal{K}_{SO} =& -2\alpha_0 (s_x \sin k_y - s_y \sin k_x)+ 2\alpha_1 \tau_x (s_x \sin k_x - s_y \sin k_y) + 2\alpha_3 \tau_z (s_x \sin k_y + s_y \sin k_x).
\end{align}
In the band basis, these terms comes out to be
\begin{align}
U \left. \mathcal{K}_{SO} \right\vert_{k_x = \pi} U^\dagger =& - 2(\alpha_0 \cos \beta_{k_y} - \alpha_3 \tau_z) s_x \sin k_y -2\tau_x (\alpha_0 \sin \beta_{k_y} + \alpha_1) s_y \sin k_y\nonumber\\
\approx& -2\left[\alpha_0 \frac{2(t-t')}{\sqrt{4(t-t')^2 + \lambda^2}}- \alpha_3 \tau_z\right]s_x \sin k_y -2\tau_x \left[\alpha_0  \frac{\lambda}{\sqrt{4(t-t')^2 + \lambda^2}}+ \alpha_1\right]s_y \sin k_y.
\end{align}

We can now see that the $\tau_z= -1$ projection of the full BdG Hamiltonian is
\begin{align}
P U \left. \mathcal{H}_{BdG} \right\vert_{k_x = \pi} U^\dagger P =& \mu_z \left[\left\{(t+t')+\frac{(t-t')^2}{\sqrt{(t-t')^2 + \lambda^2/4}}\right\}(1-\cos k_y)- \delta\mu - 2\left\{\alpha_0 \frac{(t-t')}{\sqrt{(t-t')^2 + \lambda^2/4}} + \alpha_3\right\} s_x \sin k_y\right]\nonumber\\
&- s_z h_Z + \mu_x |\Delta|,
\end{align}
where $P=(1-\tau_z)/2$ is the projection operator to the lower band; this is clearly in the same form as Eq.\eqref{EQ:1D}.

\end{widetext}

\end{document}